\documentclass[11pt,onecolumn]{article}
\usepackage{graphics,graphicx,epsfig,amsmath,color}
\usepackage[T1]{fontenc}
\usepackage[left=2cm,right=2cm,top=2cm,bottom=2cm,bindingoffset=0cm]{geometry}
\usepackage{subfig}
\usepackage{lmodern}
\usepackage{authblk}
\usepackage[english]{babel}

\begin{document}
\title{The impact of electrical couplings on the sequential bursting activity in the ensemble of inhibitory coupled Van der Pol elements}
\author[1]{T. A. Levanova \thanks{tatiana.levanova@itmm.unn.ru}}
\author[1,2]{A. O. Kazakov}
\author[1]{A. G. Korotkov}
\author[1]{G. V. Osipov}
\affil[1]{Lobachevsky State University of Nizhny Novgorod, Institute of Information Technologies, Mathematics and Mechanics, 23, Prospekt Gagarina, Nizhny Novgorod, 603950, Russia}
\affil[2]{National Research University Higher School of Economics, 25/12 Bolshaya Pecherskaya Ulitsa, Nizhny Novgorod, 603155, Russia}
\maketitle
 
\abstract{The new phenomenological model of the ensemble of three neurons with chemical (synaptic) and electrical couplings has been studied. One neuron is modeled by a single Van der Pol oscillator. The influence of the electrical coupling strength and the frequency mismatch between the elements to the regime of sequential activity is investigated.}

\section{Introduction}

In the last few years, a new field of medicine that is called bioelectronic medicine \cite{Birmingham2014} is actively developing. The main feature of bioelectronic medicine is the application of electrical stimulus to nervous system and body's tissues instead of chemical (pharmaceutical) treatment. The main target of electrical impact is nerve fibers, and signals are delivered to them using implants or wearable devices. The reasons for such interest to bioelectronic medicine are related both to the rapid improvement of technology (among the factors we can note the emergence of biocompatible soft electronics, the rapid growth in computing performance, the small size of the devices \cite{Seo2016}), and a limited success of pharmacology in the treatment of neurological disorders by medication. In this regard, it is also worth noting that in the coming decades the problem of treating diseases of the nervous system will become more relevant, because of the tendency of aging of the population and the growing stress in the modern world. For this reason, a number of scientific and industrial companies have paid attention to bioelectronic medicine. As recent papers show, \cite{Birmingham2014, Sacramento2018}, this approach can be successfully applied not only to treatment of diseases of the nervous system, but also in treatment of cardiovascular, inflammatory, metabolic and endocrine diseases, as evidenced by animal tests and clinical trials. The nervous system is the main regulator of internal processes in the body. It system affects the processes of thinking, digestion, motor activity, etc. \cite{Afr2004}. In this connection, there is an increasing interest in the study of electrical couplings in the nervous system and their role in the generation of various regimes of neuronal activity, as well as the mechanisms of their formation and suppression. The development of new medical technologies and their implementation in practical treatment requires much deeper understanding how the peripheral nervous system is involved in the regulation of various processes in the body.

The main goal of this work is to study the influence of electrical couplings to regimes of sequential bursting activity in models of neural ensembles with chemical (synaptic) couplings. For this purpose, there is considered a phenomenological model of the minimal ensemble of three non-identical neurons which demonstrate the described types of couplings. Each of the neurons is modeled by corresponding Van der Pol oscillator, but these oscillators have different proper frequencies. In our previous paper \cite{Levanova2013} the ensemble of identical Van der Pol oscillators only with chemical couplings was studied in details. In particular, there were studied various dynamical regimes occuring in this ensemble when varying the strengths of chemical couplings, and scenarios of appearance and disappearance of these regimes were also investigated. In the papers \cite{MikhaylovLevanova2013}, \cite{Levanova2016} it was shown that the obtained types of activity and the mathematical images underlying them, as well as the  scenarios of transitions from one type of activity to another are universal for a wide class of systems. In the present work we investigate the influence both of electrical couplings and the non-identity of the elements on the dynamics of the neuronal ensemble, specially focusing on the evolution of sequential bursting activity, since this regime of neural activity is very important from the point of view of neurodynamics \cite{Birds}-\cite{Insects}. We emphasise that results which are presented in the paper are quite similar, by principal qualitative properties, ro results of real biological experiments \cite{Nicholls2011}.

\section{The model}

The ensemble of three non-identical neuron-like elements connected to each other by mutual chemical (synaptic) inhibitory and electrical couplings is modeled by the following system of three Van der Pol oscillators
\begin{equation}\left\{
\begin{array}{l}
\ddot{x_j} - \mu[\lambda(x_j, \dot{x_j}) - x^2_j]\dot{x_j} + \omega_j^2x_j + d(x_{j+1}-2x_{j}+x_{j-1}) = 0,  \\
j=1,2,3.
\end{array}
\right.\label{eq1}
\end{equation}
where the variable $x_j$ phenomenologically describe the value of the membrane potential of the $j$-th neuron-like element. The electrical couplings between the ensemble elements are given by the expressions $d(x_{j+1}-2x_{j}+x_{j-1})$, where the parameter $d$ is the coefficient of the electrical coupling. The chemical (synaptic) inhibitory interaction between neuron-like elements in the ensemble is phenomenologically described in the same way as in the paper \cite{Levanova2013}, where:
\begin{equation}
\lambda(x_j, \dot{x_j})=1 - g_{1}F\left(\sqrt{x_{j+1}^2+\dot{x}_{j+1}^2}\right) - g_2F\left(\sqrt{x_{j-1}^2+\dot{x}_{j-1}^2}\right),\\
\label{eq2}
\end{equation}
\begin{figure}
	\begin{center}
		\includegraphics[width=0.5\columnwidth]{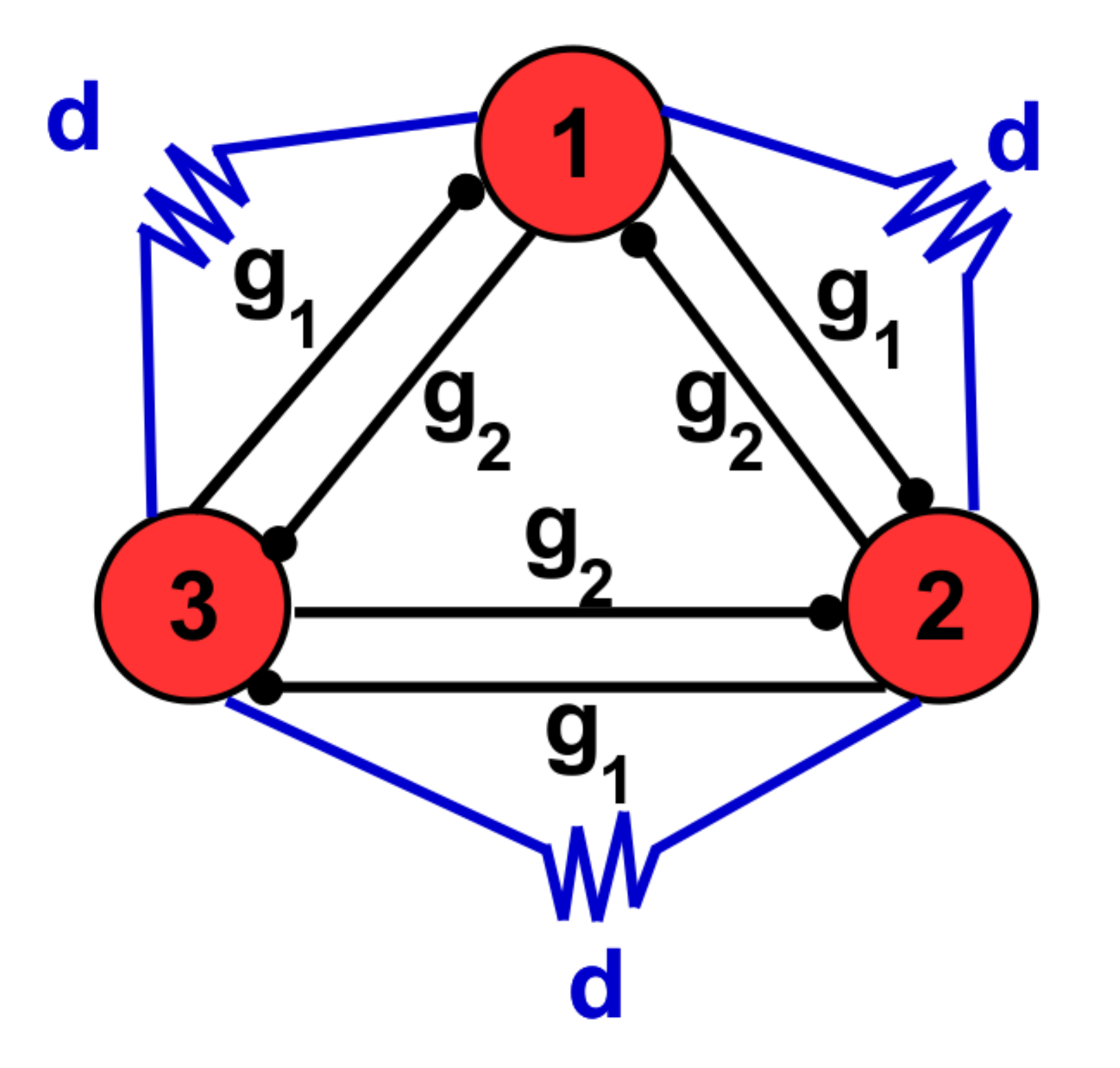}	
		\caption{The topology of the chemical (synaptic) couplings $g_1$ and $g_2$ and electrical couplings $d$ in the ensemble of neuron-like elements given by the system (\ref{eq1}).}
		\label{fig00}
	\end{center}
\end{figure}
Here $g_1$ and $g_2$ are the strengths of the inhibitory couplings directed clockwise and counter-clockwise, respectively, see Fig. \ref{fig00}. The function $F(z)$ is an activation function with a threshold value $z_0$ that phenomenologically describes the main principle of the synaptic coupling:
\begin{equation}
F(z)=\frac{1}{1+exp(-k(z-z_0))}.\\
\label{eq3}
\end{equation}
With the values $k=100$ and $z_0=0.5$ of the parameters chosen for modeling, the nonlinear function $F(z)$ is close to the step function, but remains smooth. When $z$ reaches the threshold value $z=z_0$, what corresponds to the generation of amplitude oscillations above a certain threshold one by a presynaptic element, the function $F(z)$ grows jumpwise from 0 to 1 and remains equal to 1 with a further increase of $z$. This leads, in turn, to the fact that, in the case of a sufficient coupling strength, the presynaptic neuron-like element can suppress the activity of the postsynaptic neuron-like element by generating oscillations of large amplitude. It is known that in real experiments the frequencies for different neurons and clusters of neurons are different. This allows us to introduce the parameter $\Delta$ into system (\ref{eq1}). Here $\omega_2 = \omega_1 - \Delta$, $ \omega_3 = \omega_1 + \Delta$. The parameter $\mu << 1$ determines the dynamics of a single element, in which quasi-harmonic oscillations are observed in the absence of couplings \cite{Andronov1966}.
\begin{figure}
	\begin{center}
		\includegraphics[width=0.9\columnwidth]{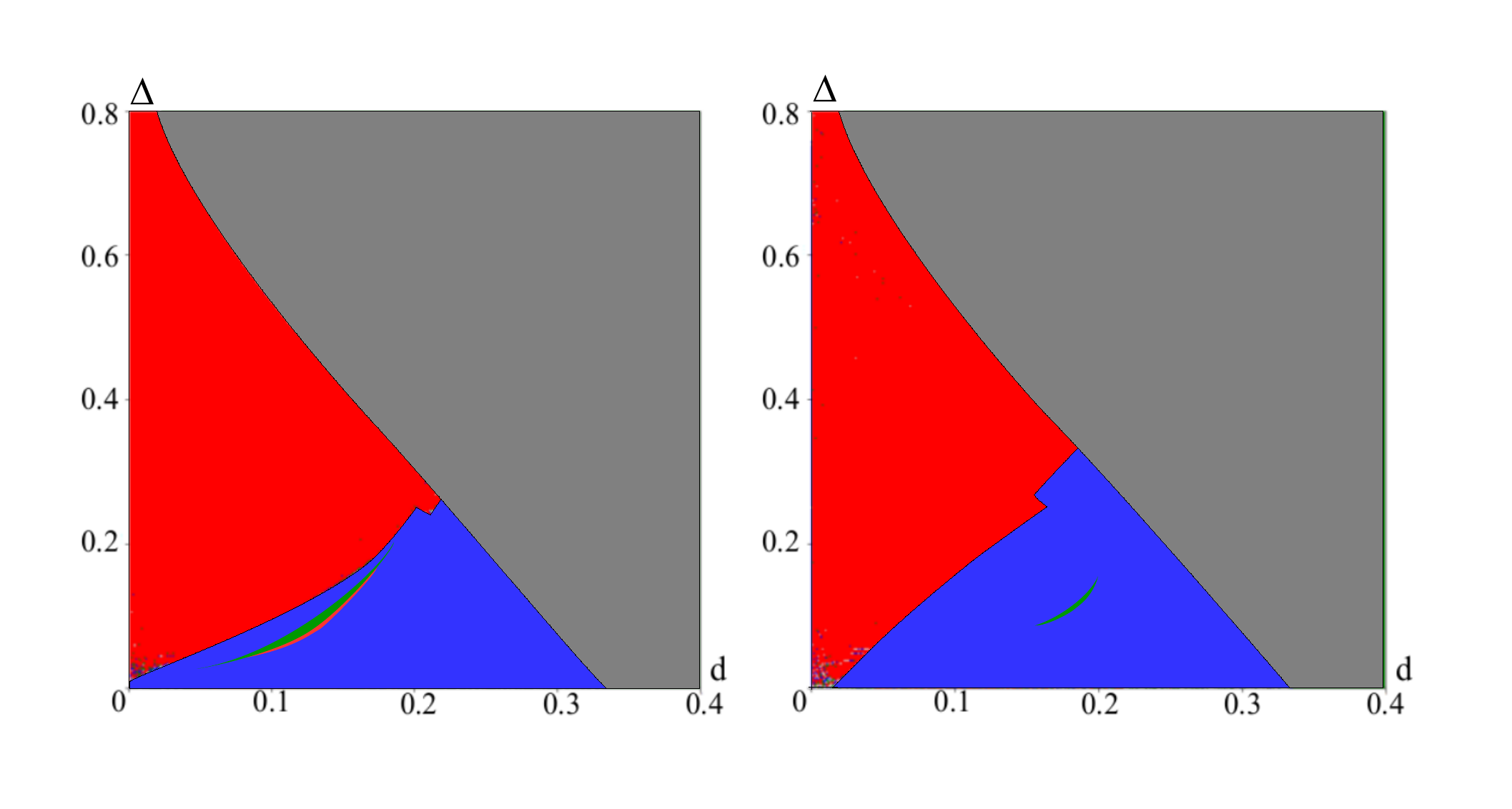}	
		\caption{Maps of the largest Lyapunov exponent of the system (\ref{eq1}). a - $(g_1, g_2) = (0,5)$; b --- $(g_1, g_2) = (5,0)$. Red points denote $\Lambda_1>0$. Green points denote $\Lambda_1=\Lambda_2=0$. Blue points denote $\Lambda_1=0$. Gray points denote $\Lambda_1<-0.0005$. The upper triangular region on the charts corresponds to the situation when the trajectories of the system go to infinity. }
			\label{fig1}
	\end{center}
\end{figure}
In \cite{Levanova2013} it was shown that the regime of sequential bursting activity is observed in system (\ref{eq1}) at $d = 0$ and $\Delta = 0$ in the case of strong asymmetry of chemical couplings. To study how the adding of electrical couplings and the frequency mismatch between the elements affects the evolution of the sequential activity regime, we built charts of the largest Lyapunov exponent on the parameter plane $(d,\Delta)$ (see Fig. \ref{fig1}). In the red color on the charts the regions corresponding to the positive largest Lyapunov exponent $\Lambda_1>0$ are marked, which indicates the presence of the chaotic dynamics in the system. Such dynamics occurs when the stable heteroclinic circuit (observed in the system at $d=0$) is destroyed by adding electrical couplings with $d\neq 0$. The absolute value of $\Lambda_1$ in this case, as a rule, is not very large. Green color on the charts denotes regions corresponding to $\Lambda_1=\Lambda_2=0$. In this case, a two-dimensional torus is observed in the phase space of the system. The regions corresponding to periodic orbits are marked with a blue and gray color. Moreover, the blue color corresponds to the regions in which the largest Lyapnov exponent fluctuates near zero, with minor deviations, of order of numerical error, the positive values ($0 <\Lambda_1 <0.0005$), and the gray color corresponds to the regions in which $-0.0005 <\Lambda_1 <0$  \footnote{This means that, for getting adequate values of Lyapunov exponents, it is required  in this case the increasing of the level of numerical accuracy. However, in both cases, as shown by numerical experiments, such regimes correspond to stable limit cycles.}. As one can see from Fig. \ref{fig1}, in both cases there is a threshold relation between $d$ and $\Delta$, above which the trajectories of the system begin to go to infinity (the upper triangular region on both charts). This fact agrees well with the data of biological experiments, which show that in real biological systems it is impossible to increase the strength of the couplings without a limitation. Within the framework of this constraint, periodic as well as quasi-periodic and chaotic regimes are observed in the system for various ratios of $d$ and $\Delta$. These regimes were not observed in the absence of an electrical couplings and the frequency mismatch.

\section{The evolution of the sequential bursting activity}

Recall that the regime of sequential bursting activity is observed in system (\ref{eq1}) for the case of strong asymmetry of the couplings \cite{Levanova2013}. For example, the value of the coupling parameter $g_1$ can be taken significantly larger than $g_2$, which is small or equal to zero. The main feature of this regime is the exponentially increasing in time the length of the burst. The mathematical image of the sequential activity regime in the phase space of the system (\ref{eq1}) is a stable heteroclinic circuit containing the saddle limit cycles. The phase orbit, which asymptotically approaching this heteroclinic circuit, spends more and more time in a vicinity of the saddle limit cycles that corresponds to the growth of the burst length.

Let us study analytically how the described heteroclinic circuit evolves in the presence of the relatively small electrical couplings and the frequency mismatch between elements. We set $\omega_1=1$, $d=\mu d_1$, $\Delta=\mu \Delta_1$. We rewrite the system of differential equations (\ref{eq1}) in the following form
\begin{equation}\left\{
\begin{array}{l}
\ddot{x}_1 + x_1= \mu[\lambda(x_1, \dot{x_1}) - x^2_1]\dot{x_1} - \mu d_1(x_2-2x_1+x_3),  \\
\ddot{x}_2 + (1+\mu \Delta_1)x_2 = \mu[\lambda(x_2, \dot{x_2}) - x^2_2]\dot{x_2} - \mu d_1(x_1-2x_2+x_3),  \\
\ddot{x}_3 + (1-\mu \Delta_1)x_3 = \mu[\lambda(x_3, \dot{x_3}) - x^2_3]\dot{x_3} - \mu d_1(x_1-2x_3+x_2).  \\
\end{array}
\right.\label{eq4}
\end{equation}
Using the Van der Pol method \cite{VanDerPol} and averaging the system by the period $T=2\pi$ we obtain the following equations for complex averaged amplitudes $z_1$, $z_2$ and $z_3$:
\begin{equation}\left\{
\begin{array}{l}
\dot{z}_1 = [\lambda(z_1, \dot{z_1}) - z_1\bar{z_1}]z_1 + id_1(z_2-2z_1+z_3),  \\
\dot{z}_2 = [\lambda(z_2, \dot{z_2}) - z_2\bar{z_2}]z_2 + id_1(z_1-2z_2+z_3) + i\Delta_1 z_2,  \\
\dot{z}_3 = [\lambda(z_3, \dot{z_3}) - z_3\bar{z_3}]z_3 + id_1(z_1-2z_3+z_2) - i\Delta_1 z_3.  \\
\end{array}\right.\label{eq5}
\end{equation}
Now introduce real amplitudes $R_1$, $R_2$, $R_3$ and phases $\phi_1$, $\phi_2$ and $\phi_3$ in the following way
\begin{equation}\left\{
\begin{array}{l}
z_1=\frac{R_1}{2}e^{-i\phi_1},\\
z_2=\frac{R_2}{2}e^{-i\phi_2}, \\
z_3=\frac{R_3}{2}e^{-i\phi_3}.\\
\end{array}\right.\label{eqz}
\end{equation}
After substituting (\ref{eqz}) in (\ref{eq5}) we obtain the following system:
\begin{equation}\left\{
\begin{array}{l}
\dot{R}_1 = [\lambda(R_1, \dot{R_1}) - \frac{R_1^2}{4}]R_1 - R_2d_1\sin(\phi_1-\phi_2) - R_3d_1\sin(\phi_1-\phi_3),  \\
\dot{R}_2 = [\lambda(R_2, \dot{R_2}) - \frac{R_2^2}{4}]R_2 - R_1d_1\sin(\phi_2-\phi_1) - R_3d_1\sin(\phi_2-\phi_3),  \\
\dot{R}_3 = [\lambda(R_3, \dot{R_3}) - \frac{R_3^2}{4}]R_3 - R_1d_1\sin(\phi_3-\phi_1) - R_2d_1\sin(\phi_3-\phi_2),  \\
R_1\dot{\phi}_1 =  2d_1R_1 - R_2d_1\cos(\phi_1-\phi_2) - R_3d_1\cos(\phi_1-\phi_3),\\
R_2\dot{\phi}_2 =  2d_1R_2 - \Delta_1R_2 - R_1d_1\cos(\phi_2-\phi_1) - R_3d_1\cos(\phi_2-\phi_3),\\
R_3\dot{\phi}_3 =  2d_1R_3 + \Delta_1R_3 - R_2d_1\cos(\phi_3-\phi_2) - R_1d_1\cos(\phi_3-\phi_1).\\
\end{array}
\right.\label{eq6}
\end{equation}
that describes the dynamics of averaged amplitudes $R_1$, $R_2$, $R_3$ and phases $\phi_1$, $\phi_2$ and $\phi_3$. In the case of absence of the electrical couplings, i.e. when $d=0$, system (\ref{eq6}) splits into two subsystems. The first subsystem contains equations for averaged amplitudes:
\begin{equation}\left\{
\begin{array}{l}
\dot{R}_1 = [\lambda(R_1, \dot{R_1}) - \frac{R_1^2}{4}]R_1,  \\
\dot{R}_2 = [\lambda(R_2, \dot{R_2}) - \frac{R_2^2}{4}]R_2,  \\
\dot{R}_3 = [\lambda(R_3, \dot{R_3}) - \frac{R_3^2}{4}]R_3.  \\
\end{array}
\right.\label{eq7}
\end{equation}
The second subsystem contains equations for phases:
\begin{equation}\left\{
\begin{array}{l}
\dot{\phi}_1 = 0,\\
\dot{\phi}_2 = - \Delta_1,\\
\dot{\phi}_3 = + \Delta_1.
\end{array}
\right.\label{eq8}
\end{equation}
The analytical study of subsystem (\ref{eq7}) was carried out before in the paper \cite{Levanova2013}. System (\ref{eq7}) was considered sequentially in the restrictions to  the invariant manifolds $R_1=0$, $R_2=0$ and $R_3=0$. In particular, for asymmetric couplings it was shown that there are three equilibria on each of the invariant planes [the unstable node $(0,0)$, the saddle $(2,0)$ and the stable node $(0,2)$]. Here the equilibria $(2, 0)$ and $(0, 2)$ are joined by a stable heteroclinic trajectory that goes from the saddle to stable node. Consequently, three heteroclinic trajectories (corresponding to phase planes $(R_1,R_2)$, $(R_1,R_3)$ and $(R_2,R_3)$ and three saddles with 2-dimension stable and 1-dimension unstable manifolds form a stable heteroclinic cycle of system (\ref{eq7}). Using equations (\ref{eq7})-(\ref{eq8}) it is easy to obtain that this result remains valid also for non-identical elements (with non-zero frequency mismatch $\Delta_1\neq 0$), however, the frequencies of the elements are different. It should be noted that heteroclinic circuits, as a rule, arise in systems with symmetry, and when the symmetry breaks down, heteroclinic circuits also disappear. In our case the non-identity of the elements breaks the symmetry, but the heteroclinic circuit still exists because the amplitude dynamics described by subsystem (\ref{eq7}) does not depend on the phases determined by equations (\ref{eq8}).

Now let us study how the adding electrical coupling changes the described above stable heteroclinic circuit.  In order to do this we will study system (\ref{eq6}) on the plane $R_2=0$:
\begin{equation}\left\{
\begin{array}{l}
\dot{R}_1 = [\lambda(R_1, \dot{R_1}) - \frac{R_1^2}{4}]R_1 - R_3d_1\sin(\phi_1-\phi_3),  \\
\dot{R}_3 = [\lambda(R_3, \dot{R_3}) - \frac{R_3^2}{4}]R_3 + R_1d_1\sin(\phi_1-\phi_3), \\
R_1\dot{\phi}_1 = 2d_1R_1 - R_3d_1\cos(\phi_1-\phi_3), \\
R_3\dot{\phi}_3 = 2d_1R_3 - \Delta_1R_3 - R_1d_1\cos(\phi_1-\phi_3), \\
R_1\sin(\phi_1-\phi_2) = R_3\sin(\phi_2-\phi_3), \\
R_1\cos(\phi_1-\phi_2) = - R_3\cos(\phi_2-\phi_3).  \\
\end{array}
\right.\label{eq9}
\end{equation}
Using two last expressions we will obtain that $\phi_1 = \phi_3$, so it is possible to rewrite system (\ref{eq9}) in the following form :
\begin{equation}\left\{
\begin{array}{l}
\dot{R_1} = [\lambda(R_1, \dot{R_1}) - \frac{R_1^2}{4}]R_1,  \\
\dot{R_3} = [\lambda(R_3, \dot{R_3}) - \frac{R_3^2}{4}]R_3, \\
R_3^2d_1=\Delta_1R_1R_3 + R_1^2d_1.  \\
\end{array}
\right.\label{eq10}
\end{equation}
\begin{figure}
	\includegraphics[width=\columnwidth]{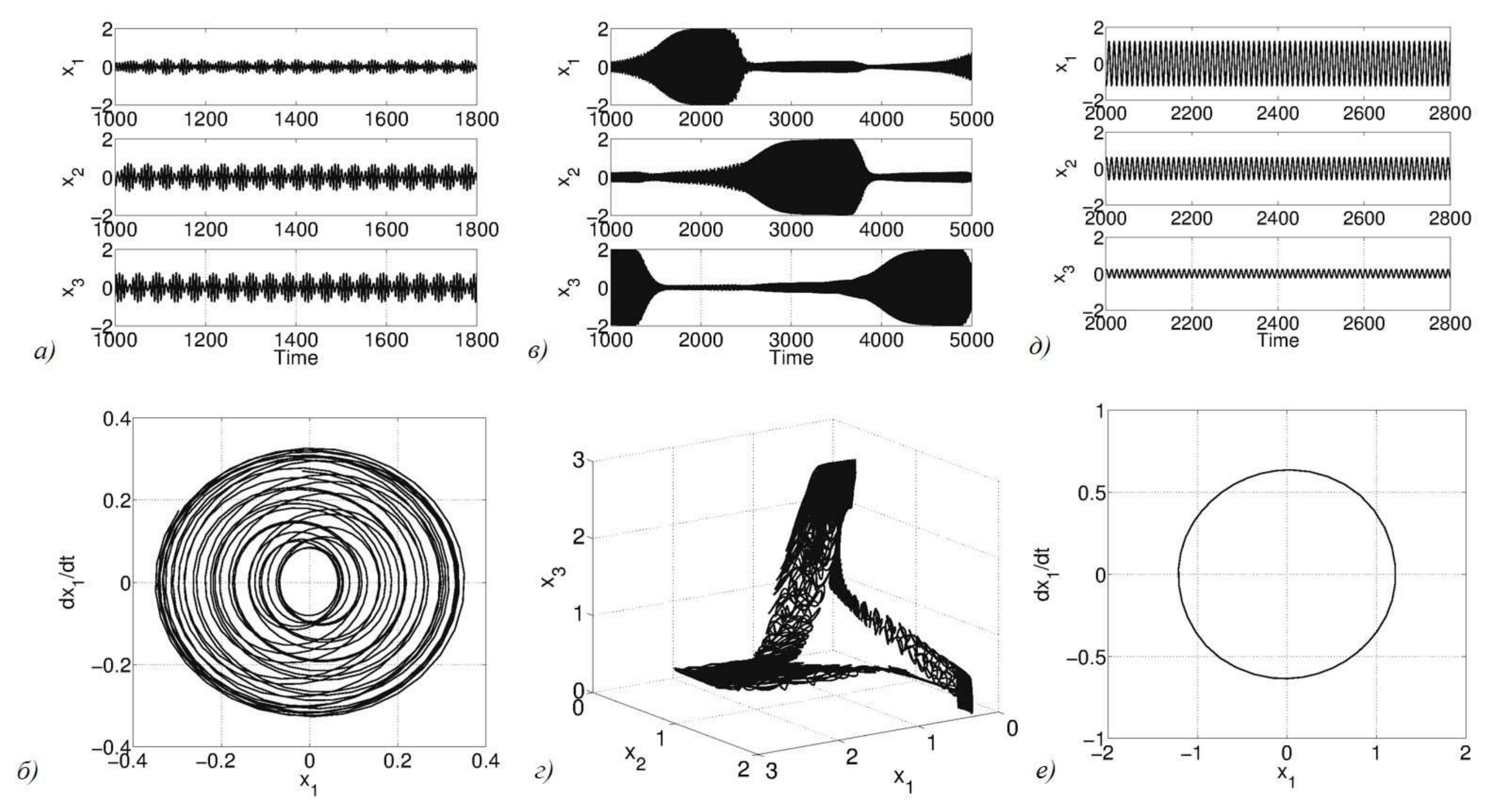}
	\caption{Time series $x_1$, $x_2$, $x_3$ and projections of the trajectories of the system (\ref{eq1}) on the 2-dimensional subspasce $(x_1, x_2)$  in the case of strong asymmetry of synaptic inhibitory couplings. Parameter values for synaptic inhibitory couplings: $(g_1, g_2)=(0,5)$. Parameter values for electrical couplings and frequency mismatch: a-b --- $d=0.1$, $\Delta=0.02$; c-d --- $d=0.1$, $\Delta=0.4$; f-g --- $d=0.2$, $\Delta=0.1$.}
	\label{fig2}
\end{figure}
It is easy to see that only the equilibria $(0,0)$ satisfy the relation from (\ref{eq10}). Two other equilibria $(2,0)$ and $(0,2)$ of system (\ref{eq7}) on the invariant manifold $R_2=0$ are no longer equilibria in the system (\ref{eq10}). Thus, introduction of a weak electrical couplings between elements leads to the fact that only the unstable equilibrium state $(0,0)$ remains in the system, and the heteroclinic circuit between saddles in system (\ref{eq6}) is destroyed. Consequently, introduction of a weak non-zero electrical couplings between elements leads to the destruction of the stable heteroclinic circuit existing in the initial system (\ref{eq1}). However, numerical experiments show that for relatively small values of $d$ (corresponding to the red points on the charts of the largest Lyapunov exponent, see Fig. \ref{fig2} a,b) in a neighborhood of the circuit there are many trajectories that sequentially visit regions near the saddle limit cycles (Fig. \ ref {fig2} b). Since the phase trajectories in this case are not attracted to the destroyed saddle limit cycles, but only visit their neighborhoods, the burst lengths for all elements are constant, and they are determined by the values of the electrical coupling and the frequency mismatch between the elements (Fig. \ref{fig2} a). The largest Lyapunov exponent is positive in this case, $\Lambda_1>0$. This scenario is similar to the scenario described in \cite{Levanova2013}, where the destruction of a heteroclinic circuit with the presence of noise was shown.

In the case when the frequency mismatch is relatively small ($0\leq \Delta \leq 0.235$), with a further increasing the strength of the electrical coupling $d$, the dynamics of the system becomes regular, namely, periodic regimes (Fig. \ref{fig2} c) and quasiperiodic oscillations (Fig. \ref{fig2} b) are observed in the numerical experiment. Upon transition to the region of quasiperiodic oscillations, a two-dimensional torus appears in the phase space of the system (Fig. \ref{fig2} d). With a further increase in the strength of the electrical coupling, the torus collapses, on its place a stable limit cycle is born (Fig. \ref{fig2} e) as a result of the supercritical Neimark-Sacker bifurcation.

\section{Conclusion}

In this paper, the ensemble of neuron-like elements is suggested to be considered as a phenomenological model of a neural network. This approach has the following advantages: it is possible to investigate low-dimensional neural models and reproduce the main effects observed in more complex models, for example, in the biologically realistic Hodgkin-Huxley model \cite{Komarov2009}, as well as in real experiments. In this study it was shown that the adding of arbitrarily small electrical couplings to the ensemble of Van der Pol neuron-like elements with chemical (synaptic) couplings leads to the destruction of a stable heteroclinic circuit between saddle limit cycles. It is also shown that the non-identity of elements (in the absence of electrical couplings) does not lead to the destruction of this heteroclinic circuit. This result is not typical for non-symmetric systems. The heteroclinic circuit exists, as a rule, in systems with symmetry, and when the symmetry breaks, it also disappears. In our case, the heteroclinic circuit is not destroyed, since the amplitude dynamics of the system does not depend on the phase one. It is also shown that after the circuit breaks down, a weak chaotic activity appears, but the further increase of the strength of the electrical couplings leads to a regularization of dynamics in the system.

These results can help to gain a deeper understanding the nature of electrical couplings in the nervous system. The study of their influence on the evolution of neuronal activity is of interest not only from the point of view of nonlinear dynamics, but also contributes to the development of the theoretical basis of bioelectronic medicine and the creation of new methods and approaches for the treatment of diseases of the nervous system that are not amenable to treatment with the help of pharmacological agents.

{\bf Acknowledgement.}

The authors thank S.V. Gonchenko for valuable advices and comments. The analytical results were obtained with the support of the RFBR grant 16-32-00835. Numerical experiments were performed within the framework of the RSF grant 14-12-00811. A. Kazakov was supported by the Basic Research Program at the National Research University Higher School of Economics (HSE) in 2018.


\begin{thebibliography}{}

\bibitem{Birmingham2014}
{Birmingham K., et al. Bioelectronic medicines: A research roadmap. Nature Reviews Drug Discover. 2014. Vol. 13. P.399.} 

\bibitem{Seo2016}
{Seo, D. et al., Wireless Recording in the Peripheral Nervous System with Ultrasonic Neural Dust, Neuron, 2016, V. 91, no 3, P. 529.}

\bibitem{Sacramento2018} 
{Sacramento J.F., et al. Bioelectronic modulation of carotid sinus nerve activity in the rat: a potential therapeutic approach for type 2 diabetes. Diabetologia. 2018. V. 61(3).  P. 700. }

\bibitem{Afr2004}
{Afraimovich V.S., et al. On the origin of reproducible sequential activity in neural circuits. Chaos. 2004. V. 14(4). P. 1123.}

\bibitem{Levanova2013}	
{Levanova T.A., et al. Sequential activity and multistability in an ensemble of coupled Van der Pol oscillators. Eur. Phys. J. Special Topics. 2013. V. 222. P. 2417.}

\bibitem{MikhaylovLevanova2013} 
{Mikhaylov A. O., et al. Sequential switching activity in ensembles of inhibitory coupled oscillators. Europhys. Lett. 2013. V. 101(2). P. 20009.}

\bibitem{Levanova2016}
{Levanova T.A., et al. Dynamics of ensemble of inhibitory coupled Rulkov maps. Eur. Phys. J. Special Topics. 2016. V. 225. P. 147.}

\bibitem{Birds}
{Fee M.S. et al. Neural mechanisms of vocal sequence generation in the songbird. Annals NY Acad. Sci. 2004. V. 1016. P. 153.}

\bibitem{Mollusk}
{Rabinovich M.I. et al. Generation and reshaping of sequences in neural systems. Biol. Cybern. 2006. V. 95. P. 519.}

\bibitem{Insects}
{Varona P. et al. Winnerless competition between sensory neurons generates chaos: A possible mechanism for molluscan hunting behavior. Chaos. 2002. V. 12. P. 672.}

\bibitem{Nicholls2011}
{Nicholls J.G., et al. From Neuron to brain. 5th ed. Sinauer Associates, 2011. 621 p.}

\bibitem{Andronov1966}
{Andronov A.A., et al. Theory of ascillations. New York: Pergamon Press, 1966.}

\bibitem{VanDerPol}
{Zwillinger D. Handbook of Differential Equations, 3rd ed. Boston: Academic Press, 1997.}

\bibitem{Komarov2009}
{Komarov M.A., et al. Sequentially activated groups in neural networks. Europhys. Lett. 2009. V. 86. P. 60006.}

\end{thebibliography}
\end{document}